\documentclass[12pt]{article}
\usepackage{epsfig,amsfonts,amssymb,amsbsy,amsbsy,array}
\def\cvp{\raise 2pt\hbox{,}}
\def\sign{\mathop{\rm sign}\nolimits}
\def\tr{\mathop{\rm tr}\nolimits}
\def\im{\mathop{\rm Im}\nolimits}
\def\re{\mathop{\rm Re}\nolimits}

\def\d{{\rm d}}

\def\plb#1#2#3{{\it Phys.\ Lett.\ }{\bf B #1} (#2) #3}
\def\npb#1#2#3{{\it Nucl.\ Phys.\ }{\bf B #1} (#2) #3}

\def\prl#1#2#3{{\it Phys.\ Rev.\ Lett.\ }{\bf #1} (#2) #3}
\def\jhep#1#2#3{{\it J. High Energy Phys.\ }{\bf #1} (#2) #3}
\def\prd#1#2#3{{\it Phys.\ Rev.\ }{\bf D #1} (#2) #3}
\def\prb#1#2#3{{\it Phys.\ Rev.\ }{\bf B #1} (#2) #3}

\def\cmp#1#2#3{{\it Comm.\ Math.\ Phys.\ }{\bf #1} (#2) #3}
\def\pr#1#2#3{{\it Phys.\ Rep.\ }{\bf #1} (#2) #3}

\def\ap#1#2#3{{\it Ann.\ of Phys.\ }{\bf #1} (#2) #3}
\begin{document}
%
%
\pagestyle{empty}
{\parskip 0in

\hfill NEIP-02-005

\hfill LPTENS-02/38

\hfill hep-th/0207066}

\vfill
\begin{center}
{\LARGE Spectral asymmetry and supersymmetry}

\vspace{0.4in}

Frank F{\scshape errari}
\\
\medskip
{\it Institut de Physique, Universit\'e de Neuch\^atel\\
rue A.-L.~Br\'eguet 1, CH-2000 Neuch\^atel, Switzerland\\ and\\
Centre National de la Recherche Scientifique\\
Laboratoire de Physique Th\'eorique de l'\'Ecole Normale 
Sup\'erieure\\
24, rue Lhomond, 75231 Paris Cedex 05, France}\\
\smallskip
{\tt frank.ferrari@unine.ch}
\end{center}
\vfill\noindent
Fractional charges, and in particular the spectral asymmetry 
$\eta$ of certain Dirac operators, can appear in the central charge of 
supersymmetric field theories. This yields unexpected analyticity 
constraints on $\eta$ from which classic results can be recovered
in an elegant way. The method could also be applied in the context of 
string theory.

\vfill

\medskip
%
\begin{flushleft}
July 2002
\end{flushleft}
\newpage\pagestyle{plain}
\baselineskip 16pt
\setcounter{footnote}{0}

\subsection*{1.~Introduction}

A field theory with fundamental fields carrying integer charges only
can have sectors in the Hilbert space with fractionally charged
states. Such states obviously cannot be created by any finite action
of the local fundamental fields. Their existence can be inferred in
the context of a semi-classical analysis \cite{JR}, where they
correspond to solitons, particle-like solutions of the classical field
equations. This important phenomenon has many applications, in
particular for polymers \cite{SSH} and the quantum Hall effect
\cite{ASW}. I recommend the excellent recent review by Wilczek
\cite{W} for more details. The purpose of the present note is to point
out some interesting properties of the fractional charges, that have
not been discussed previously in spite of the long history of the
subject. We will describe in particular an elegant way to recover the
classic results. We are also able to give some exact formulas for the
charges in some specific models, that go beyond the usual
semi-classical approximation. The ideas we will discuss can be
extended from the usual field theory setting to the more general
string theory setting, where the study of charge fractionization is
still in its infancy.

A common example of a charge that can be fractionated is the fermion
number $F$. The Dirac hamiltonian in a soliton background has in
general a non-trivial energy spectrum with a density of eigenvalues
$\rho(E)$. Semi-classically, the fermion number can be
computed in a standard way by expanding the Dirac spinor $\psi$ in
terms of positive and negative energy eigenstates. This yields
\begin{equation}
\label{Fas}
F = {1\over 2}\int\!\d x\, \langle [\psi^{\dagger},\psi] \rangle
= -{\eta \over 2}\, \cvp
\end{equation} 
where $\eta$ is the so-called spectral asymmetry \cite{APS}, the
difference between the number of positive and negative energy
eigenstates,
\begin{equation}
\label{sadef}
\eta = \lim_{h\rightarrow 0} \int\!\d E\, \rho(E)
\sign (E) |\lambda|^{-h}\, .
\end{equation}
The formulas (\ref{sadef}) and (\ref{Fas}) have respectively a mod 2
and a mod 1 ambiguity when the Dirac operator has zero modes. This is
due to degenerate lowest soliton states with different fermion
numbers. For example, when a conjugation symmetry relates states with
opposite energies, only the zero modes can contribute to the fermion
number. With $k$ complex zero modes, $F$ can then take any of the
$k+1$ different values $-k/2, -k/2+1,\ldots , +k/2$. In the most
interesting and generic cases, there is no zero mode and no
conjugation symmetry, and all the eigenvalues can contribute to $F$.
As shown in \cite{GW}, the fermion number is then in general
irrational. A detailed analysis of this problem, with many
applications and references, can be found in \cite{pr}.

A typical example is the fractional
fermion number of a magnetic monopole in an ${\rm SU}(2)$
four-dimensional Yang-Mills theory with an adjoint Higgs field. The
Dirac equation is
\begin{equation}
\label{Dmonopole}
i\gamma^{\mu} \left(\strut \partial_{\mu} + i A_{\mu}\right) \psi = 
\left( \strut m_{1} + \phi_{1} - i\gamma^{5} (m_{2}+\phi_{2})\right)
\psi\, ,
\end{equation}
where $m_{1}$ and $m_{2}$ are real mass parameters, $\phi_{1}$ and
$\phi_{2}$ are real adjoint Higgs fields, and $\psi$ is a Dirac spinor
of charge $F=1$. Asymptotically, the background fields $\phi_{j}$ tend
to the Higgs vacuum expectation values $\langle\phi_{j}\rangle =
a_{j}\sigma^{3}$. The integer magnetic number $p$ 
is given by the magnetic flux through the sphere at infinity,
\begin{equation}
\label{asymp1}
{1\over 8\pi a} \int_{{\rm S}_{\infty}}\!\! \d S^{i}\, 
\epsilon_{ijk} \tr (\phi F^{jk}) = p\, ,
\end{equation}
where $a = a_{1} + ia_{2}$ and $\phi = \phi_{1} + i\phi_{2}$.
The fermion number for this problem has been calculated in the 
literature in cases of increasing generality. In the conjugation 
symmetric case $m_{2}=a_{2}=0$, the number of zero modes $k=p$ can be 
derived using Callias' index theorem \cite{cal}. When 
$m_{1}=a_{2}=0$, the formula for $F$ was given in \cite{GW}, and when 
$a_{2}=0$ it was given in \cite{Sem},
\begin{equation}
\label{Fmon}
F = {p\over 2\pi} \left[ \arctan  \Bigl( {m_{1}-a_{1}\over 
m_{2}}\Bigr) - \arctan \Bigl( {m_{1}+a_{1}\over 
m_{2}}\Bigr) \right]\, .
\end{equation}
We will consider the more general 
Dirac operator for which $m_{1}$, $m_{2}$, $a_{1}$ and $a_{2}$ can all
be non-zero because it is the case that naturally arises in our approach. 

The formula (\ref{Fmon}) has a curious property that has not been
discussed before: \smallskip\\
{\it The fermion number or equivalently the
spectral asymmetry is a harmonic function \smallskip of the parameters.}\\
The simplest proof of this statement is given by noting that $F$ is the 
imaginary part of a holomorphic function. By using the 
complex parameters $m=m_{1}+im_{2}$ and $a=a_{1}+ia_{2}$, we have 
indeed
\begin{equation}
\label{Fhol}
F = {p\over 2\pi} \im\ln {m+a\over m-a}\, \cdotp
\end{equation}
The logarithm is defined with the branch cut on the negative real axis 
and the argument of a complex number between $-\pi$ and $\pi$. When 
$a_{2}=0$ we then recover (\ref{Fmon}), and we will prove in the next 
section that (\ref{Fhol}) is the correct generalization. The 
real part of the holomorphic function contains the terms $\ln |m\pm 
a|$, which are the logarithms of the eigenvalues of the fermion mass
matrix in (\ref{Dmonopole}).
This suggests the more precise \smallskip statement:\vfill\eject
\noindent
{\it The fermion number or equivalently the spectral 
asymmetry is given by the imaginary part of a holomorphic function 
whose real part can be deduced from the  
one-loop low \smallskip energy effective coupling of some 
field theory containing the fermion $\psi$.}\\
This result is powerful, 
because it relates a rather involved calculation of the fermion 
number in a solitonic sector to a trivial one-loop calculation in the 
vacuum sector. Moreover,
the validity of this result is not limited to the four dimensional 
Dirac operator in the monopole background. For example, in the 
two dimensional version of (\ref{Dmonopole}),
the background fields $\phi_{j}$ describe a kink solution,
$\lim_{x\rightarrow\pm\infty}\phi_{j}=\phi_{j,\pm}$,
and the vector potential 
$A_{\mu}$, that goes to a pure gauge at infinity,
implement a possible gauge symmetry. The fermion number 
has been calculated in \cite{GW,ns,pr,comtet} for $A_{\mu}$=0.
By introducing $\phi_{\pm} = 
\phi_{1,\pm} + i \phi_{2,\pm}$ it can be put in the form
\begin{equation}
\label{2dF}
F = {1\over 2\pi} \im\ln {m+\phi_{-}\over m+\phi_{+}}\,\cvp
\end{equation}
which has the same qualitative features as (\ref{Fhol}).

\subsection*{2.~Charge fractionization and supersymmetry}

The properties of $\eta$ discussed above can be checked on the final
formulas, but are rather strange and unexpected from the point of view
of the standard approach to the problem \cite{pr}. We will now present
a framework that makes those properties very natural, and from which
formulas like (\ref{Fhol}) or (\ref{2dF}) are easily derived. The idea
is to embed the problem in a supersymmetric setting. Of course
supersymmetry is not fundamental in our problem, since the objects
that we consider---the spectral asymmetry of a Dirac operator or more
generally fractional charges---are defined and mostly used in a
non-supersymmetric context. But the point is that supersymmetry is a
nice mathematical tool that provides an interesting new point of view
on those objects. The fact that the phenomenon of charge
fractionization can play an important r\^ole in the physics of
supersymmetric field theories was emphasized in \cite{fprl}. In some
sense, we will show that the arguments of \cite{fprl} can be used
backwards to infer results on charge fractionization.

The coupling of a Dirac fermion to a vector potential and a complex adjoint
Higgs field as described by (\ref{Dmonopole}) occurs in ${\cal N}=2$
supersymmetric Yang-Mills theories in four dimensions 
with one flavor of quark of complex bare mass
$m$. The Dirac fermion belongs to the quark hypermultiplet, and the
coupling to $A_{\mu}$ and $\phi$ is determined by gauge invariance and
supersymmetry. The formula of fundamental importance to us is a certain
anticommutator of the supersymmetry charges,
\begin{equation}
\label{Z}
\{ Q_{\alpha}^{I},Q_{\beta}^{J}\} = 
2\sqrt{2}\epsilon_{\alpha\beta}\epsilon^{IJ}\, Z\, ,
\end{equation}
where $\alpha$ and $\beta$ are spinorial indices and $I$ and $J$, 
$1\leq I,J\leq 2$, label the 
supersymmetry charges. The bosonic charge $Z$ is called 
the central charge of the supersymmetry algebra. Classically, it is a 
linear combination of the electric charge $Q_{\rm e}$, the magnetic 
charge $Q_{\rm m}$ and the fermion number $F$,
\begin{equation}
\label{Zcl}
Z_{\rm cl} = {2a (Q_{\rm e}+iQ_{\rm m})\over g_{\rm YM}} + mF\, ,
\end{equation}
where $g_{\rm YM}$ is the Yang-Mills couping constant. For us, the most
important property of $Z$, valid in the full quantum 
theory, is the following:
\smallskip\\
{\it The central charge $Z$
is a holomorphic function of $a$ and $m$, such that
\begin{equation}
\label{Zq}
{\partial Z\over\partial a} = p\,\tau_{\rm eff} + q\, ,
\end{equation}
where $p$ and $q$ are the integer-valued magnetic and electric quantum 
numbers respectively and
$\tau_{\rm eff}$ is the complexified low energy effective coupling
constant defined in terms of the effective Yang-Mills 
coupling and effective topological theta angle as
\begin{equation}
\label{taudef}
\tau_{\rm eff} = {\theta_{\rm eff}\over\pi} + {8i\pi\over g^{2}_{\rm 
YM,\, eff}}\,\cdotp
\end{equation}
}
This result comes from the fact that $Z$ can be calculated from the
low energy effective action \cite{SW2}, which is governed by a single
holomorphic function $\cal F$ called the prepotential such that
$\partial^{2}{\cal F}/\partial a^{2} = \tau_{\rm eff}$ \cite{sg}. The
analyticity property can also be deduced from supersymmetric Ward
identities.

The fact that the real charge
$F$ contributes to the holomorphic function $Z$ gives a 
natural explanation of the harmonicity properties discussed in section 
one. To make this idea quantitative, we need a formula expressing $F$ in 
terms of $Z$ in the full quantum theory. This is a priori non-trivial, 
because a derivation of the quantum version of (\ref{Zcl}) from the low 
energy effective action has not appear when $m\not =0$. The result, 
however, is suggested by the quantum analysis of the electric charge. 
The Witten effect \cite{weffect} in the low 
energy theory implies that
\begin{equation}
\label{We}
{2Q_{\rm e}\over g_{\rm YM,\, eff}} = q + {p\theta_{\rm eff}\over\pi}
= \re {\partial Z\over\partial a}\,\cdotp
\end{equation}
Now, (\ref{Dmonopole}) shows that the Higgs field couple to the 
electric charge in the same way as $m$ couple to the fermion number. 
We thus propose that the correct quantum formula for the fermion number
is simply
\begin{equation}
\label{Ffor}
F = \re {\partial Z\over\partial m}\,\cdotp
\end{equation}
Equations (\ref{We}) and (\ref{Ffor}) show that both $Q_{\rm e}/g_{\rm 
YM,\, eff}$ and $F$ are harmonic functions of the parameters. 
Semi-classically, the electric charge is related to a quantity,
similar to the spectral asymmetry, involving the Dirac operator
(\ref{Dmonopole}). The methods used to calculate 
$\eta$ can be straightforwardly adapted for
$a_{2}=0$ \cite{NPS}, and indeed yield a harmonic function,
\begin{equation}
\label{elec}
{2Q_{\rm e}\over g_{\rm YM,\, eff}} = q
-{p\over 2\pi} \left[ \arctan  \Bigl( {m_{1}-a_{1}\over 
m_{2}}\Bigr) + \arctan \Bigl( {m_{1}+a_{1}\over 
m_{2}}\Bigr) \right]\, .
\end{equation}

We still have to understand the relation to a one-loop effective coupling
constant. The idea is that the real parts of the derivatives of $Z$,
which are difficult to obtain directly, can be deduced from the
imaginary parts by using holomorphy. The imaginary parts turn out to be 
particularly easy to calculate. The Dirac quantization condition
\cite{Dirac} implies that
\begin{equation}
\label{Dc}
{2Q_{\rm m}\over g_{\rm YM,\, eff}} = {8\pi p\over g^{2}_{\rm YM,\, 
eff}} = \im {\partial Z\over\partial a}\,\cdotp
\end{equation}
A standard non-renormalization theorem \cite{NR} states that $g^{2}_{\rm
YM,\, eff}$ is given to all orders of perturbation theory by one-loop
Feynman diagrams, and that there is also a series of non-perturbative
contributions from instanton sectors. Let us neglect those instanton
contributions for the moment. The perturbative low energy effective
coupling is \cite{wein}
\begin{equation}
\label{effg}
\im{\partial Z\over\partial a} = {4p\over\pi} \ln {|a|\over\Lambda}
- {p\over 2\pi} \ln {|m^{2} - a^{2}|\over\Lambda^{2}}\, \cvp
\end{equation}
where we have introduced the dynamically generated scale $\Lambda$.
When $|a|\gg |m|$, the coupling is
given by the one-loop $\beta$ function of the non-abelian ${\rm
SU}(2)$ super Yang-Mills theory with one flavor of quark. When $|m|\gg
|a|$ the quark must be integrated out and the running with respect to
$a$ is given by the $\beta$ function of the pure ${\rm SU}(2)$ super
Yang-Mills theory. Around the points $a=\pm m$, the low energy theory
is an abelian gauge theory coupled to one light charged
hypermultiplet, and the infrared divergence when $a=\pm m$ is
governed by the usual infrared-free coupling of this theory.
From (\ref{effg}) we deduce
\begin{equation}
\label{tau1l}
{\partial Z\over\partial a} = q + 
{4ip\over\pi}\ln {a\over\Lambda} - {ip\over 2\pi} \ln 
{m^{2}-a^{2}\over\Lambda^{2}}\, \cvp
\end{equation}
for some integer electric number $q$. This equation can be integrated
by noting that $Z(a=0,m)=0$ because the monopole solution reduces to
the vacuum when $a=0$. The fermion number charge is then immediately
derived from (\ref{Ffor}) and we recover (\ref{Fhol}). The 
ambiguity modulo $2i\pi$ in the logarithm
is cleared up by requiring that $-p/2\leq
F\leq p/2$ in the conjugation symmetric limit. The fermion-induced
fractional electric charge of the monopole is also immediately
obtained from (\ref{tau1l}) by using (\ref{We}),
\begin{equation}
\label{Qefrac}
{2Q_{\rm e}\over g_{\rm YM,\, eff}} = q + {p\over 
2\pi}\im\ln {m^{2}-a^{2}\over a^{8}}\, \cvp
\end{equation}
in perfect agreement with (\ref{elec}) in the case $a_{2}=0$.

What about the instanton series? Equations (\ref{We}) and (\ref{Ffor})
give a precise prescription from which the exact non-perturbative
charges can in principle be calculated from the formulas of 
\cite{SW2}. This is interesting, because
to my knowledge the phenomenon of charge fractionization has never
been studied beyond the semi-classical approximation. However, the
exact charges are highly model-dependent and can be calculated only in
a supersymmetric context. On the other hand, the results of the
semi-classical approximation (\ref{Fhol}) or (\ref{Qefrac}) entirely
rely on the mathematical analysis of (\ref{Dmonopole}) and are
universal and independent of supersymmetry.

The two dimensional case with fermion number (\ref{2dF}) can be 
treated similarly. The Dirac equation (\ref{Dmonopole}) occurs in the 
coupling of a charged chiral multiplet containing the fermion $\psi$ 
with a twisted chiral multiplet containing $A_{\mu}$ and $\phi$ in a
two dimensional ${\cal N}=2$ supersymmetric gauge theory. A review on 
this type of theory can be found in \cite{ferr}. The parameter 
$m$ is often called a twisted mass in this context.
The central charge appears in the anticommutator
\begin{equation}
\label{cc2}
\{\bar Q_{+},Q_{-}\} = 4Z\, .
\end{equation}
For a soliton interpolating between two vacua $\phi_{-}$ and 
$\phi_{+}$, the classical central charge is expressed 
in terms of the tree-level twisted superpotential 
$W(\phi)$ and the fermion number,
\begin{equation}
\label{Zcl2}
Z_{\rm cl} = i\left(\strut W(\phi_{+})-W(\phi_{-})\right) + mF\, .
\end{equation}
Quantum mechanically, $Z$ is a holomorphic function of the parameters 
and is expressed in terms of an effective superpotential $W_{\rm eff}$
deduced by integrating out the charged chiral 
multiplet. This amounts to a simple one-loop calculation because the 
multiplet appears only
quadratically in the action. The result is
\begin{equation}
\label{weff}
W_{\rm eff} = {1\over 2\pi} (m+ \phi)\ln {m+\phi\over e\Lambda}\, 
\cdotp 
\end{equation}
The fermion number (\ref{2dF}) is then immediately deduced from
(\ref{Ffor}). In this case, there is no correction to the
semi-classical formula for $F$ as a function of the vacuum
expectation values $\phi_{+}$ and $\phi_{-}$ of the scalar field
$\phi$. Yet, those expectation values are model-dependent and can pick
up some non-perturbative terms.

\subsection*{3.~Prospects}

Apart from its simplicity, the most attractive feature of
our approach is that it can a priori be
generalized to string theory. String theory has solitonic states
called D-branes which are very similar to magnetic monopoles. A
detailed theory of charge fractionization for D-branes could then
certainly be developped. It would be very interesting to work out the
mathematical concepts that generalize the spectral asymmetry of the
Dirac operator which is the central object in field theory. Our method
suggests that string perturbation theory together with analyticity
constraints could be used to calculate the charges.

\subsection*{Acknowledgements}
I would like to thank Alain Comtet for discussions. This work was 
supported in part by the Swiss National Science Foundation.
\end{document}